\documentclass[12pt]{iopart}
\usepackage{graphicx}
\usepackage{dcolumn}
\usepackage{bm}
\usepackage{iopams}
\usepackage{color}

\newcommand{\ora}[1]{\textcolor[rgb]{1.0,0.5,0.0}{\bf #1}}
\def\be{\begin{equation}}
\def\ee{\end{equation}}
\def\ba{\begin{array}}
\def\ea{\end{array}}
\def\bea{\begin{eqnarray}}
\def\eea{\end{eqnarray}}
\begin{document}
\title[{}]
{Inner crust of neutron stars with mass-fitted Skyrme interaction}
\author{K. Madhuri$^{1*}$, D. N. Basu$^{2}$, T. R. Routray$^{1}$ and S. P. Pattnaik$^{1}$ }
\address{$^1$School of Physics, Sambalpur University, Jyotivihar-768019, India}
\address{$^2$Variable Energy Cyclotron Center, 1/AF Bidhan Nagar, Kolkata, 700064, India}
\ead{$^*$kmadhuriphy@gmail.com (corresponding author)}
\date{\today}

\begin{abstract}
The mass, radius and crustal fraction of moment of inertia in neutron stars are calculated using $\beta$-equilibrated nuclear matter obtained from Skyrme effective interaction. The transition density, pressure and proton fraction at the inner edge separating the liquid core from the solid crust of the neutron stars are determined from the thermodynamic stability conditions using the KDE0v1 set. The neutron star masses obtained by solving the Tolman-Oppenheimer-Volkoff equations using neutron star matter obtained from this set is able to describe highly massive compact stars $\sim$2$M_{\odot}$. The crustal fraction of the moment of inertia can be extracted from studying pulsar glitches. This fraction is highly dependent on the core-crust transition pressure and corresponding density. These results for pressure and density at core-crust transition together with the observed minimum crustal fraction of the total moment of inertia provide a limit for the radius of the Vela pulsar:  $R\geq$ 3.69+3.44 M/$M_{\odot}$. Present calculations suggest that the crustal fraction of the total moment of inertia can be at most 3.6{\%} due to crustal entrainment caused by Bragg reflection of unbound neutrons by lattice ions.


\end{abstract}



\maketitle

\bigskip

\section{Introduction}
			The study of the equation of state (EOS) of a neutron star (NS) structure is an area of contemporary research interest in nuclear astrophysics. Study of the properties of neutron stars allows us to test our knowledge on the properties of nuclear matter under extreme conditions. According to many theoretical studies, the neutron star comprises of an inner liquid core surrounded by an outer crust. The crust is made up of neutron drip out region and an inner crust \cite{sap83, bbp71, rav83}. While the neutron drip out density is relatively well determined, the transition density between inner crust to core is still not defined due to the lack of knowledge of EOS of NS matter. The core-crust interface corresponds to the phase transition between nuclei and uniform nuclear matter. The uniform matter is nearly pure neutron matter, with a proton fraction of just a few percent determined by the condition of $\beta$- equilibrium \cite{mus12}. NSs are detected from their electromagnetic radiation. These are usually observed to pulse radio waves and other electromagnetic radiation, and NSs observed with pulses are called pulsars. The identification of pulsars as rotating neutron stars has renewed interest in the properties of matter at very high densities. Pulsars are the most precise clocks with rotation periods ranging from about 1.396 milliseconds for the recently discovered pulsar J1748-2446ad \cite{hess2007}  up to several seconds. The periodicity of arrival time of pulses is extremely stable. The slight delays associated with the spin-down of the star are at most of a few tens of microseconds per year. Nevertheless, long term monitoring of pulsars has revealed irregularities in their rotational frequencies. The first kind of irregularity, called timing noise, is random fluctuations of pulse arrival times and is present mainly in young pulsars, such as the Crab, for which the slow down rate is larger than for older pulsars. Indeed, correlations have been found between the spin-down rate and the noise amplitude \cite{lyne1998}. Timing noise might result from irregular transfers of angular momentum between the crust and the liquid interior of neutron stars. A second kind of irregularity is the sudden jumps or "glitches" of the rotational frequency, which have been observed in radio pulsars and more recently in anomalous X-ray pulsars \cite{kaspi2000}. Glitch is a sudden increase in the rotational frequency of pulsar due to emission of radiation and high energy particles. This sudden increase in rotational frequency is due to the small time coupling of the inner crust of pulsar with the superfluid liquid core. The transfer of angular momentum due to the core-crust coupling decreases the measured time period of pulsar. The inner crust consists of a crystal lattice of nuclei immersed in a neutron superfluid \cite{blink1999} where core to crust transition occurs. The sudden decrease in the angular momentum of the superfluid within the crust causing a sudden increase in angular momentum of the rigid crust results into a pulsar glitch. However, almost all the theoretical phenomena agree that there is rigid structure within NS and it acquires a considerable volume which contributes a significant amount in the total moment of inertia of NS structure.
		
		In the present work skyrme force, KDE0v1 is used for the reason disscussed in sub-section (2.3). 
		The core-crust transition density and the corresponding pressure is determined \cite{datta2014} by considering the $\beta$- stability condition of  nuclear matter with respect to the thermodynamic stability conditions \cite{jml2007,kubis2004,kubis2007,worley2008,callen1985}. The Tolman-Oppenheimer-Volkoff Equation (TOV) \cite{tol1939,opp1939} is used to derive the mass and radius relationship of NS. The crustal fraction of total moment of inertia is determined by using core-crust transition density and corresponding pressure \cite{blink1999} which allows us to limit the radius of vela pulsar.

\section{Formalism}
\label{Sec:formalism}
The core of the neutron star is dominantly composed of $\beta-$equilibrated dense $ n+p+e+\mu$ matter in liquid phase and matter as a whole is charge neutral. The conditions for $\beta-$equilibrium on charge neutrality are \cite{bayma}
\begin{eqnarray}
\mu_{n}-\mu_{p}=\mu_{e}=\mu_{\mu},
\label{eq1}
  \end{eqnarray}
  \begin{eqnarray}
Y_{p}=Y_{e}+Y_{n},
\label{eq2}
  \end{eqnarray}
  where $\mu_{i}$ and $ Y_{i}$, i=n, p, e, $\mu$ are the respective chemical potentials and particle fractions. Here n, p, e, $\mu$ are neutron, proton, electron and muon respectively.
  
  The transition density is calculated from the onset of instability of the uniform liquid against small amplitude density fluctuation due to clusterization. The basic requirement of the thermodynamic method is that the system must satisfy the stability conditions $V_{thermal}>0$ which is given by \cite {kubis2007,cai12}
\begin{eqnarray}
V_{thermal}= 2\rho \left(\frac{\partial\ e(\rho,Y_{p})} {\partial\rho} \right) + 
\rho^{2} \left (\frac {\partial^{2} e(\rho,Y_{p})}{\partial\rho^{2}}\right)
-\frac{\left(\rho \frac{\partial^{2}e(\rho,Y_{p})}{\partial\rho \partial Y_{p}}\right)^{2}}
{\left ({\frac{\partial^{2}e(\rho,Y_{p})}{\partial Y_{p}^{2}}}\right)}>0 
\label{eq3}
\end{eqnarray}
where $V_{thermal}= \left(\frac{\partial P^{N}}{\partial \rho}\right)_{\mu}$,(where $P{^N}$ is the baryonic pressure) and $e(\rho,Y_{p} )=\frac{H(\rho,Y_{p})}{\rho}$
is the energy per baryon, $H(\rho,Y_{p})$ being the energy density. The transition density $\rho_t$, calculated from the stability condition  given by Eq.(\ref{eq3})  \cite{mous10,datta2014,trr2016} by setting $V_{thermal}=0$, plays an important role in crustal fraction of moment of inertia. The crustal fraction of moment of inertia is important for the explanation of the observed glitches in the pulsars \cite{trr2016}.

\subsection{\it Crustal fraction of the moment of inertia of neutron stars}
The crustal fraction of moment of inertia $\Delta {I}/{I}$ contains the mass $M$ and  radius $R$ of the 
NS and is given by the following approximate expression \cite {xu09,link99}
\begin{eqnarray} 
\frac {\Delta{I}}{I} &\approx \frac{28\pi P({\rho_t})R^{3}}{3Mc^{2}} \left( \frac{1-1.67\xi-0.6\xi^{2}}{\xi} \right)\nonumber\\
&\times \left(1+\frac{2P({\rho_t})}{\rho_{t}mc^{2}} \frac{(1+7\xi)(1-2\xi)}{\xi^{2}} \right)^{-1} 
 \label{eq4}
  \end{eqnarray}
where $\xi=\frac {GM}{Rc^2}$, $G$ is the gravitational constant and $c$ is the velocity of light. From Eq.(\ref{eq4}) it is clear that the crustal fraction of moment of inertia has major dependence on the pressure at the crust-core transition, $P(\rho_{t})$, since 
the transition density $\rho_{t}$ enters only as correction term. The crustal fraction of moment of inertia can be inferred from the observations of pulsar glitches \cite{datta2014}. As per reference \cite{link99} glitches represent a self-regulating instability for which the star prepares over a waiting time and the glitches in the Vela pulsar suggest that the angular momentum should be such that more than 1.4{\%} of the moment of inertia drives these events. Hence, it can be inferred that if glitches originate in the liquid of the inner crust, it would imply that $\Delta {I}/{I}$ $>1.4 \%$

\subsection{\it Tolman-Oppenheimer-Volkoff equation and mass – radius relation}
The NS mass and radius for a given EOS are calculated by solving the 
Tolman-Oppenheimer-Volkov (TOV) equations numerically using Runge-Kutta method. The TOV equation is given by \cite{TOV39a,TOV39b} 

\vspace{-0.0cm}
\begin{eqnarray}
\frac{dP(r)}{dr} = -\frac{G}{c^4}\frac{[\varepsilon(r)+P(r)][m(r)c^2+4\pi r^3P(r)]}{r^2[1-\frac{2Gm(r)}{rc^2}]} \\ 
{\rm where} ~\varepsilon(r)=(\epsilon + m_b c^2)\rho(r),~m(r)c^2=\int_0^r \varepsilon(r') d^3r' \nonumber
\label{eq5}
\end{eqnarray}
\noindent 
where $m(r)$, $\varepsilon(r)$ and $P(r)$ are the mass of the star contained inside radius $r$, the energy density and pressure at a radial distance $r$ from the centre, respectively. 

%
The total moment of inertia, $I$, of the neutron star is calculated using the definition $I=J/{\Omega}$ by assuming the star to be rotating slowly such that its angular velocity $\Omega$ is uniform, where the expression of J is given by \cite {arn77}
\begin{eqnarray}
J=\frac{c^{2}}{6G} R^{4} \frac{d \bar {\omega}}{dr}\Big|_{r=R} 
\label{eq6}
 \end{eqnarray}
Where, $\bar {\omega}$ is the angular velocity of a point in the star measured with respect to the angular velocity of the local inertial frame.
\subsection {Skyrme interaction}{\ora {\textbf{}}
The Skyrme interaction used in the present study is given by  \cite {dutra12}:
\begin{eqnarray}
V(\rho_{n},\rho_{p})&=\frac{t_0}{4} \big[ (2+x_0)\rho^{2}-(1+2x_{0})(\rho_{n}^{2}+\rho_{p}^{2})\big]\\\nonumber
&+\frac{t_3}{24}\rho^{\gamma} \big[ (2+x_3)\rho^{2}-(1+2x_{3})(\rho_{n}^{2}+\rho_{p}^{2})\big]\\\nonumber
&+\frac{1}{8}\rho^{\gamma} \big[ t_{2}(1+2x_{2})-t_{1}(1+2x_{1})\big](\tau_{n} \rho_{n}+\tau_{p} \rho_{p})\\\nonumber
&+\frac{1}{8}\rho^{\gamma} \big[ t_{1}(2+x_{1})+t_{2}(2+x_{2})\big]\tau \rho\\\nonumber
\label{eq7}
\end{eqnarray}
where, $\rho_n+\rho_p=\rho= $total nucleon density, $\tau_i=\frac{3}{5}k_{i}^{2} \rho_{i}$, $\tau=\tau_{p}+\tau_{n}$ and $i=n,p$.
  The Skyrme interaction contains altogether nine parameters, namely, $\gamma$, $t_0$, $x_0$, $t_1$, $x_1$, $t_2$, $x_2$, $t_3$, $x_3$.
There are 240 Skyrme interaction parameter sets \cite{dutra12}, out of which only 16 parameter sets namely, GSkI, GSkII, KDE0v1, LNS, MSL0, NRAPR, Ska25s20, Ska35s20, SKRA, SkT1, SkT2, SkT3, Skxs20, SQMC650, SQMC700, SV-sym32 satisfy a series of criteria derived from macroscopic properties of nuclear matter in the vicinity of nuclear saturation density at zero temperature and their density dependence, derived by the liquid drop model, experiments with giant resonances and heavy-ion collisions. Additional constraints like density dependence of the neutron and proton effective mass  Landau parameters of symmetric and pure neutron nuclear matter,beta-equilibrium matter, and observational data on high- and low-mass cold neutron stars further reduce this number to five (KDE0v1, LNS, NRAPR, SKRA, and SQMC700, collectively known as the CSkP set) {\cite{dutra12}. Further, more constraints like maximum mass and the corresponding central density of high-mass neutron stars restricts the skyrme model to describe the NS structure because it requires extrapolation to densities above the valid range. None of the CSkP models produces a maximum mass neutron star model with central density in line with observation except KDE0v1 parameter set. Particularly KDE0v1 set tackles all the above constraints and can explain NS structure. Therefore, KDE0v1 set has been used in the present text. 

\begin{table}
\centering
\caption{Values of the nine parameters of ANM for the Skyrme interaction corresponding to KDE0v1 \cite {dutra12}}
\renewcommand{\tabcolsep}{0.05cm}
\renewcommand{\arraystretch}{1.2}
\begin{tabular}{|c|c|c|c|c|c|c|c|c|c|c|c|}\hline
\hline
$\gamma$ & $t_0$ & $x_0$ & $t_1$ & $x_1$ & $t_2$ & $x_2$ & $t_3$ & $x_3$\\\hline
0.17&-2553.1 & 0.65&411.7&-0.35 &-419.9&-0.93&14603.6&0.95 \\\hline
\multicolumn{9}{|c|}{Nuclear matter properties at saturation density} \\
\hline
\multicolumn{1}{|c|}{$\gamma$}&\multicolumn{1}{|c|}{$\rho_0$ ($\mathrm{fm}^{-3}$)} & \multicolumn{2}{c|}{$e (\rho_0) $ (MeV)}
& \multicolumn{1}{c|}{$K (\rho_0)$ (MeV)} & \multicolumn{1}{c|}{$\frac{m^*}{m}(\rho_0,k_{f_0})$}
& \multicolumn{1}{c|}{$E_s (\rho_0)$ (MeV)} & \multicolumn{2}{c|}{$L (\rho_0)$ (MeV)} \\
\hline
\multicolumn{1}{|c|}{0.17}& \multicolumn{1}{|c|}{0.165} & \multicolumn{2}{c|}{-16.0} & \multicolumn{1}{c|}{227.54}
& \multicolumn{1}{c|}{0.74} & \multicolumn{1}{c|}{34.58} & \multicolumn{2}{c|}{54.69} \\\hline
\end{tabular}
\end{table}
\begin{figure}
\vspace{0.6cm}
\begin{center}
\includegraphics[width=0.8\columnwidth,angle=-90]{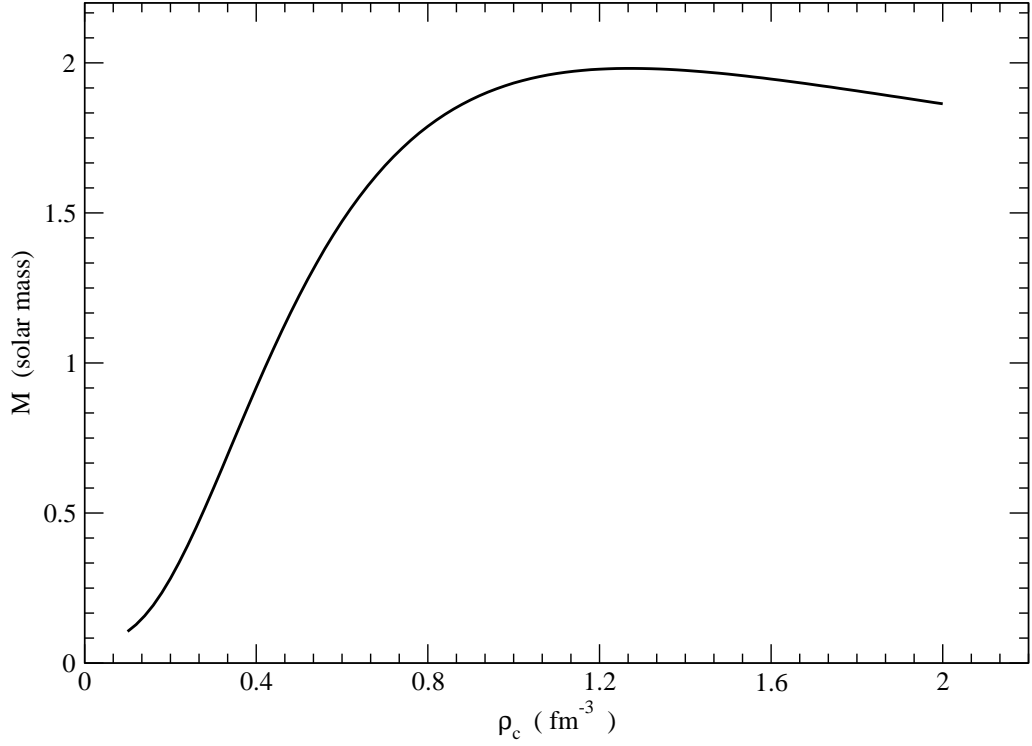}
\caption{Variation of mass with central density for slowly rotating
neutron stars for the present nuclear EOS}
\label{Figure.1}
\end{center}
\end{figure}
\begin{figure}
\vspace{0.6cm}
\begin{center}
\includegraphics[width=0.8\columnwidth,angle=-90]{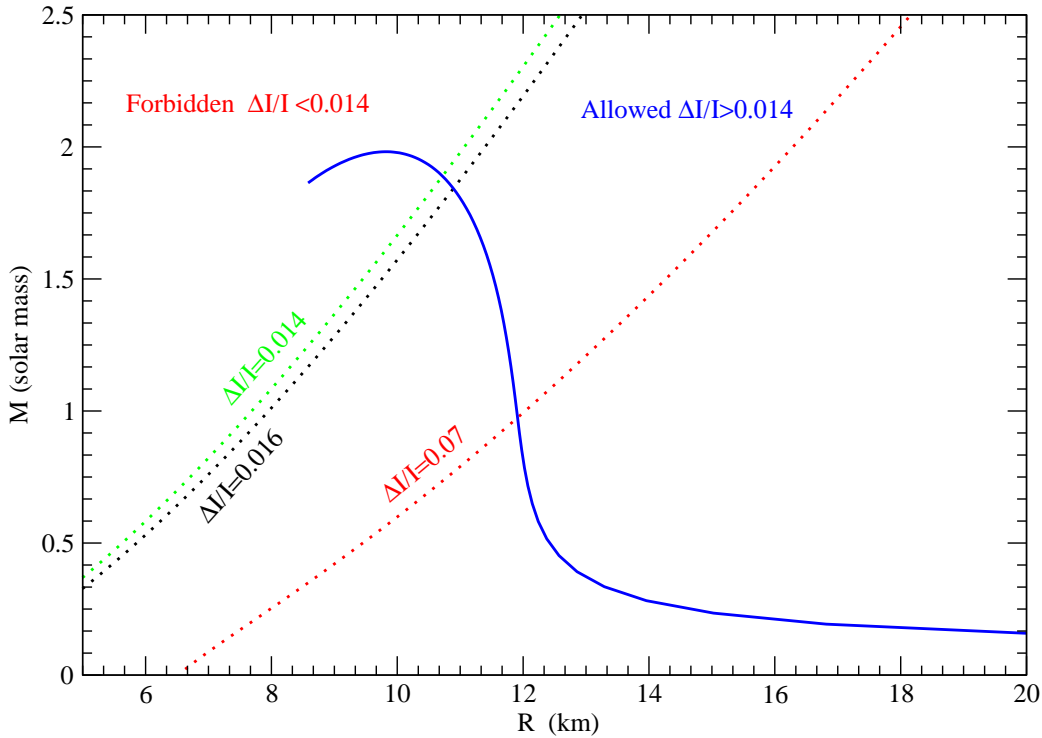}
\caption{The mass-radius relation of slowly rotating neutron stars for the present nuclear EOS. The constraint of $\frac{\Delta I}{I} > 1.4\%~(1.6\%,7\%)$ for the Vela pulsar implies that to the right of the line defined by $\frac{\Delta I}{I} = 0.014~(0.016,0.07)$ (for $\rho_t=$ 0.0904 fm$^{-3}$, P$_t=$ 0.5013 MeV fm$^{-3}$) }
\label{Figure.2}
\end{center}
\end{figure}
\begin{table}
\centering
\caption{Neutron star radius , mass ,total and crustal fraction of the moment of inertia  and crustal thickness as functions of the central density $\rho_{c}$ of the star.}
\renewcommand{\tabcolsep}{0.12cm}
\renewcommand{\arraystretch}{0.88}
\begin{tabular}{|c c c c c c||c c c c c c|}
\hline
$\rho_{c}$&$R$&	$M$& $I$ &$\frac{\Delta I}{I}$&$\delta R$ &$\rho_{c}$&$R$&	$M$& $I$ &$\frac{\Delta I}{I}$&$\delta R$\\
$\mathrm{fm}^{-3}$&km&	$M_{\odot}$ & $M_{\odot}$ $\mathrm{km}^{2}$&fraction&km&$\mathrm{fm}^{-3}$&km&	$M_{\odot}$ & $M_{\odot}$ $\mathrm{km}^{2}$&fraction&km\\\hline
 2.00&   8.5835&   1.8636&    73.11&     .0050&           .2330  &1.10& 10.2342&   1.9646&    108.11&    .0102&           .3756 \\
 1.98&   8.6079&   1.8679&    73.74&     .0050&           .2341  &1.08& 10.2854&   1.9598&    108.71&    .0105&           .3838 \\
 1.96&   8.6328&   1.8723&    74.37&     .0050&           .2350  &1.06& 10.3374&   1.9543&    109.25&    .0108&           .3925\\
 1.94&   8.6581&   1.8766&    75.02&     .0051&           .2361  &1.04& 10.3897&   1.9480&    109.76&    .0112&           .4015 \\
 1.92&   8.6839&   1.8810&    75.68&     .0051&           .2372  &1.02& 10.4426&   1.9408&    110.19&    .0116&           .4113\\
 1.90&   8.7102&   1.8853&    76.35&     .0051&           .2384  &1.00& 10.4962&   1.9328&    110.55&   .0120&           .4217\\
 1.88&   8.7370&   1.8896&    77.02&     .0052&           .2397  & .98& 10.5500&   1.9238&    110.86&    .0124&           .4325 \\
 1.86&   8.7643&   1.8939&    77.71&     .0052&           .2410  & .96& 10.6043&   1.9138&    111.07&    .0129&           .4441 \\
 1.84&   8.7920&   1.8982&    78.41&     .0053&           .2423  & .94& 10.6591&   1.9027&    111.20&    .0134&           .4566\\
 1.82&   8.8202&   1.9024&    79.12&     .0053&           .2437  & .92& 10.7141&   1.8905&    111.25&    .0139&           .4698\\
 1.80&   8.8490&   1.9067&    79.85&     .0054&           .2453  &.90&  10.7694&   1.8770&    111.18&    .0145&           .4838\\
 1.78&   8.8782&   1.9109&    80.58&     .0054&           .2467  &.88&  10.8248&   1.8623&    111.00&    .0152&           .4988\\
 1.76&   8.9081&   1.9150&    81.32&     .0055&           .2484  &.86&  10.8804&   1.8461&    110.71&    .0158&           .5149\\
 1.74&   8.9384&   1.9192&    82.07&     .0056&           .2501  &.84&  10.9361&   1.8286&    110.28&    .0166&           .5322 \\
 1.72&   8.9693&   1.9232&    82.83&     .0056&           .2518  &.82&  10.9917&   1.8094&    109.71&    .0174&           .5506 \\
 1.70&   9.0007&   1.9272&    83.60&     .0057&           .2530  &.80&  11.0472&   1.7887&    109.01&    .0182&           .5705\\ 
 1.68&   9.0327&   1.9312&    84.39&     .0058&           .2556  &.78&  11.1023&   1.7662&    108.12&    .0192&           .5919 \\
 1.66&   9.0652&   1.9351&    85.18&     .0058&           .2575  &.76&  11.1573&   1.7420&    107.07&    .0202&           .6150 \\
 1.64&   9.0983&   1.9389&    85.98&     .0059&           .2596  &.74&  11.2116&   1.7158&    105.87&    .0213&           .6399\\
 1.62&   9.1320&   1.9426&    86.80&     .0060&           .2618  &.72&  11.2651&   1.6876&    104.45&    .0226&           .6666 \\
 1.60&   9.1663&   1.9462&    87.62&     .0061&           .2641  &.70&  11.3181&   1.6574&    102.78&    .0239&           .6958\\
 1.58&   9.2012&   1.9497&    88.44&     .0062&           .2665  &.68&  11.3701&   1.6250&    100.93&    .0254&           .7276 \\
 1.56&   9.2367&   1.9531&    89.28&     .0062&           .2689  &.66&  11.4211&   1.5903&    98.87&     .0270&           .7621\\
 1.54&   9.2728&   1.9564&    90.12&     .0063&           .2715  &.64&  11.4708&   1.5533&    96.58&     .0288&           .7998\\
 1.52&   9.3096&   1.9596&    90.97&     .0064&           .2743  &.62&  11.5189&   1.5139&    94.08&     .0308&           .8408\\
 1.50&   9.3470&   1.9626&    91.83&     .0065&           .2771  &.60&  11.5657&   1.4720&    91.32&     .0330&           .8860\\
 1.48&   9.3850&   1.9654&    92.69&     .0066&           .2802  &.58&  11.6106&   1.4276&    88.34&     .0355&           .9356\\
 1.46&   9.4236&   1.9681&    93.55&     .0068&           .2832  &.56&  11.6536&   1.3806&    85.12&     .0382&           .9903\\
 1.44&   9.4630&   1.9706&    94.42&     .0069&           .2865  &.54&  11.6948&   1.3311&    81.67&     .0412&          1.0510 \\
 1.42&   9.5029&   1.9729&    95.29&     .0070&           .2900  &.52&  11.7340&   1.2791&    77.99&     .0446&          1.1183\\
 1.40&   9.5436&   1.9750&    96.16&     .0071&          .2936   &.50&  11.7714&   1.2245&    74.11&     .0484&          1.1936 \\
 1.38&   9.5849&   1.9768&    97.04&     .0073&           .2973  &.48&  11.8069&   1.1675&    70.03&     .0527&          1.2778\\
 1.36&   9.6268&   1.9784&    97.91&     .0074&           .3013  &.46&  11.8409&   1.1082&    65.78&     .0575&          1.3727\\
 1.34&   9.6695&   1.9797&    98.78&     .0076&           .3055  &.44&  11.8739&   1.0467&    61.39&     .0630&          1.4801 \\
 1.32&   9.7128&   1.9807&    99.64&     .0077&           .3098  &.42&  11.9070&    .9833&    56.90&     .0691&          1.6025\\
 1.30&   9.7569&   1.9814&    100.50&    .0079&           .3143  &.40&  11.9412&    .9182&    52.34&     .0760&          1.7428\\
 1.28&   9.8016&   1.9817&    101.34&    .0081&           .3191  &.38&  11.9787&    .8518&    47.76&     .0839&          1.9050 \\
 1.26&   9.8470&   1.9817&    102.17&    .0082&           .3243  &.36&  12.0220&    .7844&    43.22&     .0928&          2.0939 \\
 1.24&   9.8931&   1.9813&    103.00&    .0084&           .3296 &.34&  12.0758&    .7166&    38.77&     .1030&          2.3165\\
 1.22&   9.9398&   1.9804&    103.81&    .0086&           .3351 &.32&  12.1458&    .6488&    34.48&     .1146&          2.5814\\
 1.20&   9.9872&   1.9791&    104.59&    .0089&           .3410 &.30&  12.2412&    .5817&    30.42&     .1279&          2.9009\\
 1.18&  10.0354&   1.9774&    105.35&    .0091&           .3472 &.28&  12.3759&    .5160&    26.65&     .1433&          3.2925\\
 1.16&  10.0841&   1.9750&    106.10&    .0093&           .3537  &.26&  12.5702&    .4524&    23.25&     .1616&          3.7815\\
 1.14&  10.1335&   1.9722&    106.80&    .0096&           .3607  &.24&  12.8573&    .3916&    20.29&     .1840&          4.4061\\
 1.12&  10.1835&   1.9687&    107.48&    .0099&           .3679  &.22&  13.2898&    .3346&    17.83&     .2125&          5.2264\\\hline

\end{tabular}
\end{table}
\begin{table}
\centering
\caption{The straight line equation for the radius $R$ of the Vela pulsar as a function of its mass $M$ in the exact calculation for the EOS of table 1.}
\renewcommand{\tabcolsep}{0.0991cm}
\renewcommand{\arraystretch}{1.2}
\begin{tabular}{|c|c|c|c|}\hline\hline
 $\rho_{t}$          &  $ P_{t}$             & $Y_{p}(\rho_{t})$  &    Vela Pulsar \\
		 			           & 					             &  					        &   Radius constraint\\\hline
$\mathrm{fm}^{-3 }$  & MeV$\mathrm{fm}^{-3}$ &                    &   km\\\hline

0.0904               &0.5013	               &0.0415              &$R\geq$ 3.69+3.44M/$M_{\odot}$ \\\hline

\end{tabular}
\end{table}

\section{Results and Discussion} 
\label{Sec:res}
NSs	with masses more than 1.8$M_{\odot}$ known as radio pulsars, are important probes of nuclear astrophysics in extreme conditions. They posses extreme internal gravitational fields, which results in gravitational binding energies substantially higher than those found in common, 1.4$M_{\odot}$ NSs. In ref \cite{blink1999}, it was shown that vela pulsar glitches are thought to occur due to transfer of angular momentum between inner crust and the superfluid core, could be explained if at least 1.4{\%} of the total moment of inertia of the NS resides in the inner crust. Large pulsar frequency glitches can be interpreted as sudden transfer of angular momentum between the neutron superfluid permeating the inner crust and the rest of the star. In spite of the absence of viscous drag, the neutron superfluid is strongly coupled to the crust due to non-dissipative entrainment effects. It is often argued that these effects may put a constraint on the maximum  amount of angular momentum that during glitches can possibly be transferred \cite{cham2013}. We found that the EOS which is used in our present work incorporate large crustal moment of inertia and that large enough transition pressure can be generated to explain the large Vela pulsar glitches without invoking an additional angular-momentum reservoir beyond that confined to the solid crust.
		The core-crust transition density is calculated by using thermodynamical spinodal method which gives good agreement with the other two methods used to calculate transition density and pressure namely, dynamical spinodal within the Vlasov formalism and the relativistic random phase approximation \cite{agrawal2016, Ducoin2008, cjh2008}. It should be noted here that the crustal region of the compact star in the present work consists of FMT+BPS+BBP up to number density of  0.0582 $\mathrm{fm}^{-3}$ and $\beta$-equilibrated neutron star matter up to core-crust transition number density of 0.0904 $\mathrm{fm}^{-3}$ which is far beyond 0.0582 $\mathrm{fm}^{-3}$, otherwise unified EOS would have been taken. The mass-radius relationship in neutron stars is obtained by solving the TOV equations. The maximum NS mass for the EOS obtained using the KDE0v1 Skyrme set is $\sim$ 1.98 M$_\odot$ with corresponding radius of $\sim$ 9.85 km. It is important to mention here that recent observations of the binary millisecond pulsar J1614-2230 by P. B. Demorest et al. \cite{De10} suggest that the masses lie within 1.97$\pm$0.04 M$_\odot$ where M$_\odot$ is the solar mass. Recently the radio timing measurements of the pulsar PSR J0348 + 0432 and its white dwarf companion have confirmed the mass of the pulsar to be in the range 1.97-2.05 M$_\odot$ at 68.27$\%$ or 1.90-2.18 M$_\odot$ at 99.73$\%$ confidence \cite{An13}. Radii, masses, total and crustal fraction of moment of inertia and crustal thickness as function of central density  $\rho_{c}$ are listed in tabe 2. From table 2, it is clear that, pulsars with masses 1.8$M_{\odot}$ or less have crustal fraction of total moment of inertia greater than 0.014 i.e $\frac{\Delta I}{I}$ $>0.014$. The alowed mass and radius for vela pulsar is found out to be $R\geq$3.69+3.44M/$M_{\odot}$ and suggest that the crustal fraction of moment of inertia can be at most 3.6{\%} due to crustal entrainment.

\section{Summary and conclusions}
\label{Sec:con}
The mass-radius relation in vela pulsar is obtained by calculating the crustal fraction of moment of inertia in neutron star. The calculation of crustal fraction of moment of inertia requires the transition density and pressure at the transition density apart from the bulk properties of the neutron star. The transition density and pressure is calculated in the thermodynamical spinoidal formulation. These calculations are made using the Skyrme KDE0v1 parameter set which has been found to give maximum NS mass in conformity with the maximum mass constraint, apart from the several other nuclear matter constraints as discussed in sub-section 2.3. The neutron star core-crust transition density, pressure and proton fraction at transition density are found to be $\rho_{t}$=0.0904  $\mathrm{fm}^{-3}$, P$_t=$ 0.5013 MeV $\mathrm{fm}^{-3}$ and $Y_{p}(\rho_{t})$= 0.0415 respectively. The extracted value of the crustal fraction of total moment of inertia, $\frac{\Delta I}{I}$ $>1.4\%$, from the observed glitches in vela-pulsar allows us to limit the radius of vela pulsar to $R\geq$3.69+3.44M/$M_{\odot}$. The calculation also suggest, as can be seen from table 2, that the crustal fraction of total moment of inertia can be as large as 3.6{\%} due to crustal entrainment which is in agreement with Simple Effective Interaction \cite{trr2016} and Density Dependent M3Y interaction \cite{at2016}.
%

{\it {}}
 
%
%
\section*{References}

\end{document}